\documentclass[pss]{wiley2sp} 
\usepackage{amsmath}
\usepackage{amssymb,amsmath}
\usepackage{bbold}
\usepackage{graphicx}
\begin{document}

\title{Numerical method for  disordered quantum phase transitions in the large$-N$ limit}

\titlerunning{Numerical method for  disordered  phase transitions}

\author{%
 David Nozadze\textsuperscript{\Ast},
  Thomas Vojta}

\authorrunning{D. Nozadze, T. Vojta}

\mail{e-mail
  \textsf{dn9z2@mst.edu},}

\institute{%
 Department of Physics, Missouri University of Science $\&$ Technology, Rolla, MO 65409, USA}

\received{XXXX, revised XXXX, accepted XXXX} 
\published{XXXX} 


\abstract{%
%
%
%
\abstcol{%
We develop  an efficient numerical method to  study the quantum critical behavior of disordered systems with $\mathcal{O}(N)$ order-parameter symmetry in the large$-N$ limit. It is based on the iterative solution of the large$-N$  saddle-point
equations combined with a fast algorithm for inverting the arising large sparse random matrices. As an example,
we consider the superconductor-metal quantum phase transition in disordered nanowires.  
We study the behavior of various observables near the  quantum phase transition.
Our results agree with recent renormalization group predictions, i.e., the  transition is}{ governed by an infinite-randomness critical point,  accompanied by quantum Griffiths
singularities.  In contrast to the existing numerical approach to this problem, our method gives direct access to the temperature
dependencies of observables. Moreover, our algorithm is highly efficient because
the numerical effort for each iteration scales linearly with the system size. This allows us to study larger systems, with
up to 1024 sites,  than
previous methods.  We also discuss generalizations to higher dimensions  and other systems  including the itinerant antiferromagnetic transitions in disordered metals.  
 }}

%
%

\maketitle   

\section{Introduction}
\label{sec:intro12}

Randomness can have much more dramatic effects at quantum phase transitions than at classical phase transitions because quenched 
disorder is perfectly correlated in the imaginary time direction which needs to be included at quantum phase transitions. Imaginary time acts as an additional coordinate  with infinite extension
at absolute zero temperature.  Therefore, the impurities and defects are effectively very large  which  leads to 
strong-disorder phenomena including power-law quantum Griffiths singularities \cite{1969_Griffiths_PRL,1995_Thill_Physyca_A,1996_Rieger_PRB}, infinite-randomness critical points  characterized by exponential
scaling  \cite{1992_Fisher_PRL,1995_Fisher_PRB}, and smeared phase transitions \cite{2003_Vojta_PRL}. For example, 
the zero-temperature quantum phase transition in the random transverse-field Ising model  is governed by an infinite-randomness critical point \cite{1995_Fisher_PRB} featuring
slow {\it{activated}} (exponential)  rather than power-law dynamical scaling. It is accompanied by  quantum Griffiths singularities.  This means, observables are expected to be singular not
just at criticality  but in a whole parameter region near the quantum critical point which is called the quantum Griffiths phase.

Quantum Griffiths singularities are caused by rare spatial configurations of the disorder.
Due to statistical fluctuations, one can always find  spatial regions (rare regions)   which are impurity free.   
The probability $\mathcal{P}(V_{\rm{RR}})$ to find such a  rare region is exponentially small in its volume $V_{\rm{RR}}$, $\mathcal{P}(V_{\rm{RR}})\sim \exp (-bV_{\rm{RR}})$ with
$b$  being a constant that depends on the disorder strength.
Close to a magnetic phase transition, the rare region can be locally in the magnetic phase while the bulk system is still non-magnetic. When the characteristic energy $\epsilon$ of such a rare region decays exponentially with its volume, $\epsilon\sim\exp(-c V_{\rm{RR}} )$  (as in the case of the transverse-field Ising model), the resulting    rare-region density of states has power-law form,   $\rho(\epsilon) \sim \epsilon^{\lambda-1}$, where $\lambda=b/c$ is the non-universal
Griffiths exponent.  $\lambda$ takes the value zero at the quantum critical point and increases throughout the quantum Griffiths phase. The singular density of states of the rare regions leads to
 quantum Griffiths singularities of several thermodynamic observables including order-parameter susceptibility,  $\chi\sim T^{\lambda -1}$,   specific heat, $C\sim T^\lambda,$  entropy, $S\sim T^\lambda,$ and zero-temperature magnetization-field curve $m \sim h^{\lambda}$ (for reviews see, e.g., Refs. \cite{2006_Vojta_JPhysA,2010_Vojta_JLTPhys}).

Many interesting models in statistical mechanics and field theory contain 
some integer-valued parameter $N$ and can be solved  in the large$-N$ limit. Therefore, the large$-N$ method
is a very useful tool to study classical and quantum phase transitions. An early example is the Berlin-Kac spherical model \cite{1952_Berlin_RP} which is equivalent to a classical $\mathcal{O}(N)$
order parameter field theory in the large$-N$ limit \cite{1968_Stanley_PRB}.  Analogously, the  quantum spherical model \cite{1994_Vojta_PRB,1994_Nieuwenhuizen_PRL,1996_Vojta_PRB}  has been used to
investigate quantum critical behavior. In both cases, $N$ is the number of order parameter components. Another potential application of the large$-N$ method are  $\mathcal{SU}(N)$ Kondo models \cite{1987_Bickers_RMP} with spin-degeneracy $N$.
 In all of these cases, the partition 
function can be evaluated in saddle point approximation in the limit $N \gg 1$, leading to 
self-consistent equations. In  clean systems, these
equations   can often be solved analytically. However, in the presence of disorder, one obtains a
large number of coupled self-consistent equations which can  be solved only numerically.  

In this paper, we develop a new  efficient numerical method to study  critical behavior of disordered system with $\mathcal{O}(N)$ order-parameter symmetry in the large$-N$ limit.  In contrast to the existing numerical approach to this problem \cite{2008_Maestro_PRL}, our method gives direct access to the temperature
dependencies of observables. We apply this method  to the superconductor-metal quantum phase transition in disordered nanowires. Using a strong-disorder renormalization group, it has recently been predicted that this transition is in the same universality class as the random transverse-field Ising model.    We confirm  these predictions numerically. We   find the behaviors of observables as a function of temperature and an external field. They  follow the expected  quantum Griffiths power laws.  We consider up to 3000 disorder realizations   for system sizes $L=256$ and  1024. The paper is organized as follows: In Sec.~\ref{sec:model12} we introduce  the
model: a continuum Landau-Ginzburg-Wilson  order-parameter field theory in the presence of dissipation; and we generalize the theory to quenched disordered systems.  Then, we discuss the predicted critical behavior of this model and  derive the large$-N$ formulation. 
 In Sec.~\ref{sec:three_Adr}, we review an existing numerical approach to  this model.   In Sec.~\ref{sec:ourmethod}, we present our  numerical  method to study  the quantum critical behavior.   We discuss the results   in Sec.~\ref{sec:results12},  and we compare them to the behavior  predicted by the strong-disorder renormalization group. 
 Sec.~\ref{sec:perf} is devoted to  the computational performance of our method. Finally, we  conclude in Sec.~~\ref{sec:concl12} by discussing and comparing  our numerical method to the existing one. We also discuss generalizations to higher dimensions and other models.

\section{The model}
\label{sec:model12}

We start from the quantum Landau-Ginzburg-Wilson free-energy functional for an $N-$component
vector order parameter $\varphi$ in one space dimension.  For a
clean system with overdamped order parameter dynamics the Landau-Ginzburg-Wilson action reads,\footnote{We set Planck's constant and Boltzmann constant to unity ($\hbar=k_{B}=1$) in what follows.}

\begin{align} \label{action1}
S=&\frac{1}{2}\int dx \int_0^{1/T}d\tau   \Bigl[  \alpha \varphi^2(x,\tau )
+ J[\partial_x\varphi(x,\tau )]^2  \nonumber\\
& +\frac{u}{2N}  \varphi^4(x,\tau )\Bigr]+\frac{\gamma T}{2} \sum_{\omega_n} |\omega_n|\int dx |\tilde{\varphi}(x,\omega_n)|^2 \nonumber\\
&  - h \int dx \int_0^{1/T}d\tau \varphi(x,\tau )\,, 
\end{align}
where $\alpha$  is the bare distance from criticality.  $\gamma$ and  $J$  are   the strength of dissipation  and  interaction, respectively.
$u$ is the standard quartic coefficient.  $h$ is a uniform external field conjugate to the order parameter. $\tilde{\varphi}(x,\omega_n)$ is the Fourier transform of the order parameter ${\phi}(x,\tau)$ with respect to imaginary time, and  $\omega_n=2\pi n T$ is a Matsubara frequency.
The above action  with $N=2$ order parameter components
(equivalent to one complex order parameter) has been used to describe \cite{2004_Sachdev_PRL}  the superconductor-metal transition in nanowires \cite{2006_Rogachev_PRL}. 
This 
transition is driven by pair-braking interactions, possibly due to
random magnetic moments trapped on the wire surface \cite{2006_Rogachev_PRL}, which also introduce quenched disorder in the nanowire. The action (\ref{action1}) can be generalized to $d=3$ space dimensions   and $N=3$ order parameter components, in this case, it
describes itinerant antiferromagnetic quantum phase transitions \cite{1976_Hertz_PRB,1993_Millis_PRB}.

In the presence of quenched disorder, the functional form of Eq.~(\ref{action1}) does not change qualitatively.
However, the coupling constants  become    random functions of position $x$.  The full effect of disorder
can be realized by setting   $u=\gamma=1$ while considering the couplings $\alpha$ and $J$ to be randomly distributed  in  space \cite{1971_Tucker_PRB}.
The quantum phase transition in  zero external field can be tuned by changing the mean of the  $\alpha_i$ distribution, $\overline{\alpha}$.

Recently, the model (\ref{action1}) has been investigated by means of a strong-disorder renormalization  group method \cite{2007_Hoyos_PRL,2009_Vojta_PRB} (for a review of the method, see, e.g., \cite{2005_Igloi_PR}). This theory predicts that the model falls in the same universality class as the one-dimensional random transverse-field Ising model which was studied extensively by Fisher \cite{1995_Fisher_PRB}. Thus, the phase transition is characterized by an infinite-randomness critical point at which the dynamical scaling is exponential instead of power-law. Off criticality, the behaviors of
observables are characterized by strong quantum Griffiths singularities.

Let us focus on the Griffiths phase on the disordered side of
the transition, where the distance from quantum criticality 
$\delta=\bar{\alpha}-\bar{\alpha}_c>0$.  The strong-disorder renormalization group predicts the
 disorder averaged equal-time correlation function $C(x)$ to behave as
\cite{1995_Fisher_PRB}
\begin{align} \label{corrf0}
C(x)\sim  \frac{\exp[-(x/\xi)-(27\pi^2/4)^{1/3} (x/\xi)^{1/3}]}{(x/\xi)^{5/6}} \,
\end{align}
for large distances $x$. Here, $\xi$ is the correlation length which diverges as $\xi\sim |\delta|^{-\nu}$ with $\nu=2$ as  the quantum critical point is approached. The disorder averaged order parameter as a function of the external field $h$ in the Griffiths phase   has the singular form \cite{1995_Fisher_PRB}
\begin{align} \label{Orderpgrif}
\varphi(h)\sim h^{\lambda}\,.
\end{align}
Here, $\lambda$ is the non-universal Griffiths exponent which vanishes at quantum criticality as $\lambda \sim \delta^{\nu \psi}$ with critical exponent $\psi=1/2$.
Right at  criticality, the theory predicts logarithmic 
behavior rather than a power law \cite{1995_Fisher_PRB},
\begin{align} \label{Orderpcrit}
\varphi(h)\sim {[\log(h_0/h)]^{\phi-1/\psi}} \,.
\end{align}
Here, the exponent $\phi=(1+\sqrt{5})/2$ equals to the golden mean, and $h_0$ is some microscopic field scale.

The average order parameter susceptibility as a function of temperature $T$  in the disordered   Griffiths phase is expected to have the form \cite{1995_Fisher_PRB}
\begin{align} \label{Ordsusc0}
\chi(T)\sim T^{\lambda-1} \,
\end{align}
with the same $\lambda-$exponent as in Eq.~(\ref{Orderpgrif}).

Our goal is to test the strong-disorder renormalization group predictions by means of a numerical method. As a first step, we  discretize the continuum model (\ref{action1}) in space and  Fourier-transform from imaginary time $\tau$ to Matsubara frequency $\omega_n$. 
The discretized Landau-Ginzburg-Wilson   action has the form

\begin{align} \label{action3}
S=&\frac{T}{2} \sum_{i=1}^{L}\sum_{\omega_n}  \Bigl[  \alpha_i |\tilde{\varphi}_i(\omega_n)|^2
+ J_{i}|\tilde{\varphi}_i(\omega_n)-\tilde{\varphi}_{i+1}(\omega_n)|^2  \nonumber\\
&+\frac{1}{2N}  |\tilde{\varphi}_i(\omega_n)|^4\Bigr] + \sum_{i=1}^{L}\Bigl[\frac{ T}{2}\sum_{\omega_n}  |\omega_n||\tilde{\varphi}_i(\omega_n)|^2  \nonumber\\
&- h \tilde{\varphi}_i(0)\Bigr]\,, 
\end{align}
where $L$ is the system size. The nearest-neighbor interactions $J_{i}>0$ and the mass terms $\alpha_i$ (bare local distances from quantum criticality) are random quantities. 
The critical behavior of the model (\ref{action3}) can be studied in the limit of a
large number of order parameter components $N$.  In this limit, the above action can be reduced to  a Gaussian form. This can be done in several ways, for example by decomposing the square of each component of the order parameter $|\tilde{\varphi}^{(k)}_i(\omega_n)|^2$ into its average $\langle \varphi^2\rangle$ and  fluctuation $\Delta| \tilde{\varphi}^{(k)}_i(\omega_n)|^2$: $|\tilde{\varphi}^{(k)}_i(\omega_n)|^2=\langle \varphi^2\rangle +\Delta| \tilde{\varphi}^{(k)}_i(\omega_n)|^2$.  Substituting this  into the quartic term of the action (\ref{action3}) and  using the central limit theorem,  the quartic term  can be replaced by $\langle \varphi^2\rangle |\tilde{\varphi}_i(\omega_n)|^2$.   This leads to the
Gaussian action

\begin{align} \label{action2}
S=&\frac{T}{2}\sum_{i,j=1}^{L}\sum_{\omega_n} \tilde{\varphi}^{\ast}_j(\omega_n)(M_{ij}+|\omega_n|\delta_{i,j}) \tilde{\varphi}_j(\omega_n)\nonumber \\ &+h \sum_{i=1}^{L} \tilde{\varphi}_i(0) \,.
\end{align}
The coupling matrix is given by

\begin{align} \label{cplmatr}
M_{ij}=-J_i\delta_{i,j+1}-J_j\delta_{i,j-1}+(r_i+2J_i)\delta_{i,j}  \,.
\end{align}
The renormalized local distance $r_i$  from criticality at site $i$ must be determined self-consistently from
\begin{align} \label{rend}
r_i= \alpha_i + \langle \varphi_i^2\rangle\,,
\end{align}
where $\langle \varphi_i^2\rangle $ is given by 
\begin{align} \label{avgso}
\langle\varphi_i^2\rangle=T\sum_{\omega_n}[M+|\omega_n|\mathbb{1}]^{-1}_{ii}+h^2\sum_{j,k=1}^{L} M^{-1}_{ij}M^{-1}_{ik}\,.
\end{align}
Here,  $\mathbb{1}$ is the identity matrix. In the presence of disorder, the self-consistent
equations (\ref{rend}) at different sites are not identical. We thus arrive at a large number of coupled non-linear self-consistent   equations. Therefore,  numerical techniques are required to solve them.

\section{Existing numerical approach} 
\label{sec:three_Adr}

In this section, we review  the numerical method proposed by Del Maestro {\it{et al.}}  \cite{2008_Maestro_PRL}  to  study  the model  (\ref{action2})  at  zero temperature and in the absence of an external field ($h=0$).
  The matrix $M$ is spectral decomposed in terms of its orthogonal eigenvectors $V_{ij}$ and eigenvalues  $\epsilon_i$   as 
\begin{align} \label{decom}
\sum_{j=1}^{L}M_{ij}V_{jk}=V_{ik}\epsilon_k\,.
\end{align}

 Using this decomposition,  the inverse   matrix   in Eq.~(\ref{avgso}) can be written as 
 \begin{align} \label{inv21}
[M+|\omega_n|\mathbb{1}]^{-1}_{ij}=\sum_{k=1}^{L}\frac{V_{ik}V_{kj}}{\epsilon_k+|\omega_n|}\,.
 \end{align}
 
At zero temperature the sum over Matsubara frequencies in Eq.~(\ref{avgso}) turns into an integral which can be performed analytically. This  leads to the self-consistent  equations (for $h=0$),
\begin{align} \label{adsel}
\frac{1}{\pi}\sum_{j=1}^{L}(V_{ij})^2\log\left(1+\frac{\Lambda_{\omega}}{\epsilon_j}\right)+\alpha_i-r_i=0\,.
\end{align}
Here, for convergence of the frequency integral, an ultra violet cutoff $\Lambda_{\omega}$ is introduced.   
Numerical solutions to  Eq.~(\ref{adsel}) were 
 obtained by an iteration process using a modified Powell's hybrid method.  The method works well for large distances from criticality and
small system sizes, but it becomes computationally prohibitive  near criticality where the correlation length $\xi$ becomes of order of the system size.
This problem  can be partially overcome by implementing a clever iterative solve-join-patch procedure. However, the system size $L$ is still limited
because  large matrices need to be fully diagonalized  which requires  $ \mathcal{O} (L^3)$  operations per iteration.
Therefore, for  large $L$ the  method gets very slow.

As the result,  the largest sizes studied in Ref.~\cite{2008_Maestro_PRL} were $L=128$.  The authors analyzed    equal time correlations, energy gap statistics and dynamical susceptibilities and found them in agreement with the strong-disorder renormalization group predictions \cite{2007_Hoyos_PRL,2009_Vojta_PRB}. The method was also used  in Ref.~\cite{2010_Maestro_PRL} to study the conductivity.

\section{Method}
\label{sec:ourmethod}

We now present a novel numerical method to study  the model (\ref{action2}) 
  at non-zero temperatures. Its numerical effort scales linearly with system size $L$ (per iteration) compared with the $L^3$
  scaling of the numerical method outlined in Sec.~\ref{sec:three_Adr}.
   The basic idea of our method is that, for $h=0$, we only need the diagonal elements of the inverse matrix  $[M+2\pi n T\mathbb{1}]^{-1}$ to iterate the self-consistent Eq.~(\ref{rend}). The numerical effort for 
finding the diagonal elements of the inverse of a sparse matrix is much smaller than that of a full diagonalization.   Combining Eqs.~(\ref{rend}) and (\ref{avgso}), the system of self-consistent equations   at non-zero temperatures $T$, and  in the presence  of an external field $h$, reads

\begin{align} \label{self12}
r_i=2 T\sum_{n=1}^{m}Y_{ii}+T M^{-1}_{ii}+h^2\sum_{j,k=1}^{L} M^{-1}_{ij}M^{-1}_{ik} +\alpha_i \,,
\end{align}
where
\begin{align}
Y_{ii}=[M+2\pi n T\mathbb{1}]^{-1}_{ii} \,, \nonumber
\end{align}
and  $m=\Lambda_{\omega} (2\pi T)^{-1}$ with an ultra-violet cutoff  frequency $\Lambda_{\omega}$.
To solve these equations (\ref{self12}) iteratively, we find the   inverses of the tridiagonal\footnote{We use open boundary conditions.} matrices  $[M+2\pi n T \mathbb{1}]$   and $M$ using the fast method proposed in Ref.~\cite{1992_Meurant_SJMAA}. This algorithm is summarized in Appendix  \ref{app:12}.
In  zero external field, we only need the diagonal elements of $[M+2\pi n T\mathbb{1}]^{-1}$ and the number of operations per iteration  scales linearly with system size $L$, while
it scales quadratically in the presence of a field because for $h\neq 0$, full inversion of the matrix  $M$ is required.

Once the full set of $r_i$ has been obtained, we can compute observables from the quadratic action  (\ref{action2}).  Let us first consider observables in the absence of an external field. The  equal-time correlation  function $C(x) = \overline{\langle \varphi_x(\tau) \varphi_1(\tau)  \rangle}$  averaged over disorder realizations can be obtained from Eq. (\ref{action2}),

\begin{align} \label{corrf1}
C(x)=\frac{T}{L-x}\overline{\sum_{i=1}^{L-x}\left(\sum_{n=1}^{m}2Y_{i,i+x}+M^{-1}_{i,i+x}\right)} \,,
\end{align}
where the overbar  indicates  the average over disorder configurations.
Similarly, in the zero  external field, we can calculate the order parameter susceptibility as a function of temperature.  The disorder-averaged order parameter susceptibility $\chi(T)$ can be expressed as
\begin{align} \label{susc1}
\chi(T)=\frac{T}{L}\overline{\sum_{i=1}^L\sum_{k=1}^L M^{-1}_{i k}}\,.
\end{align}

In the presence of an external field, we need to include $h$ in 
the solution of Eq.~(\ref{self12}).   We can then   compute the order-parameter ${\it{vs.}}$ field curve.
The disorder-averaged order parameter reads
\begin{align} \label{orert}
\varphi(h)=   \frac{h}{L}\overline{\sum_{i=1}^L\sum_{k=1}^L M^{-1}_{i k}}\,.
\end{align}

We note that the number of operations to calculate observables for one disorder configuration  scales quadratically with the system size $L$. However, this needs to be done only once, outside the loop
that iterates the self-consistent equations.   At low temperatures, according to Eq.~(\ref{self12}), we need to  invert a huge number of matrices $[M+2\pi n T \mathbb{1}]$ per iteration (one for each Matsubara frequency). Naively, one might therefore expect   the numerical effort  to scale linearly in $1/T$.  However, these matrices are not very different. We can therefore 
accelerate the method by combining them appropriately. This is explained in Appendix \ref{app:11}.

\begin{figure}[t]
\centering
\includegraphics[width=8.2cm]{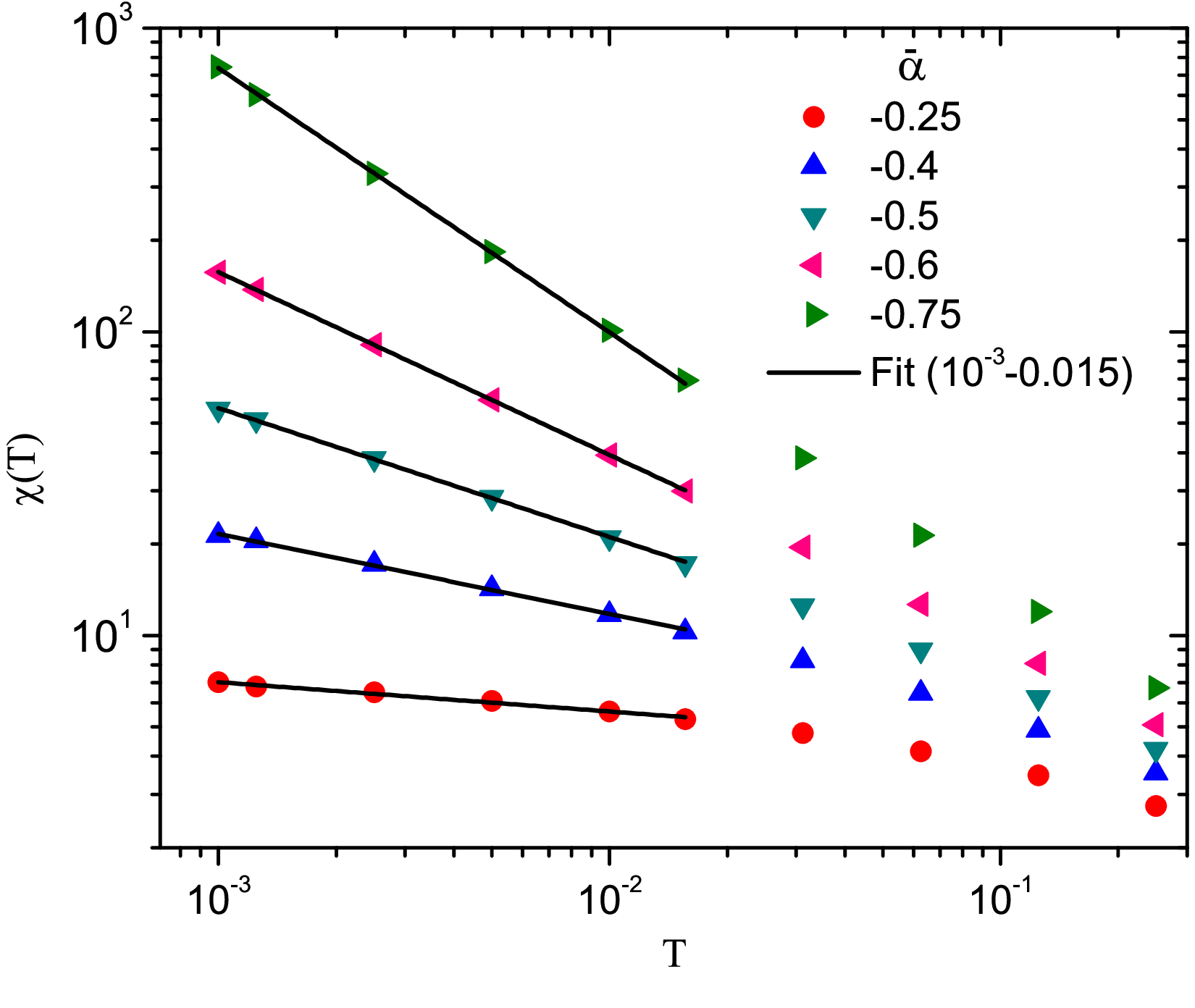}
\caption{(Color online) Order-parameter susceptibility $\chi$ versus temperature $T$ for various  $\bar{\alpha}$  in the Griffiths phase. All data are averaged
over 3000 disorder configurations with system size  $L=256$. The solid lines represent fits to the Griffiths power law            
 (\ref{Ordsusc0}), $\chi(T)\sim T^{\lambda -1}$, over the temperature range $T=10^{-3}-1.5\times 10^{-2}$.}
\label{fig:fig2}
\end{figure}

\section{Results}
\label{sec:results12}

In this section, we report results of our numerical calculations of the   model (\ref{action2}).
We consider the interactions $J_{i}$ to be uniformly distributed on $(0,1)$ with  mean $\overline{J}=0.5$ and the bare  local  distances from quantum criticality $\alpha_i$ to be
Gaussian distributed with mean $\overline{\alpha}$ and variance 0.25.

\begin{figure}[t]
\centering
\includegraphics[width=9cm]{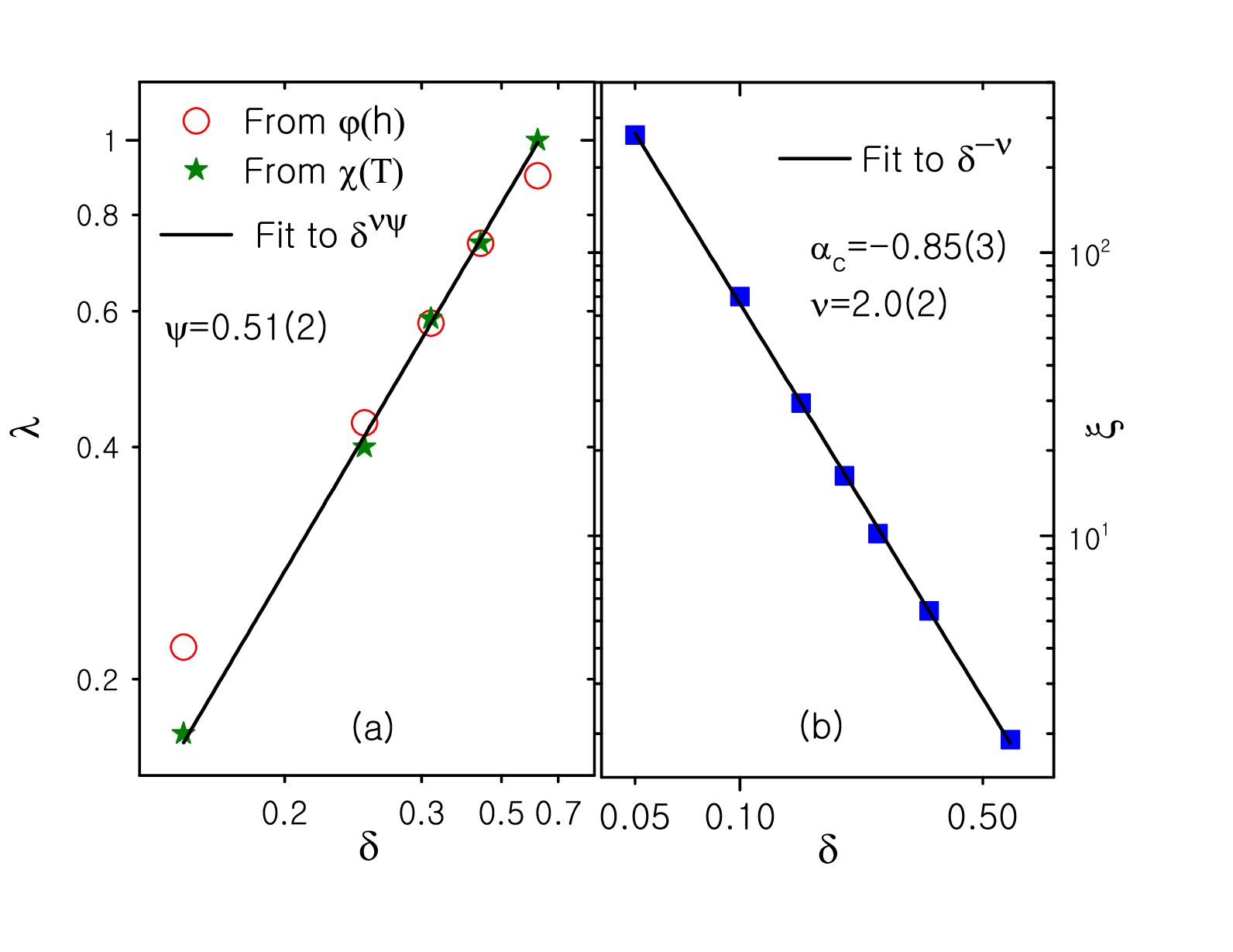}
\caption{(Color online) a) The Griffiths exponent $\lambda$ versus distance from quantum criticality $\delta$. The solid line is a fit to the power law $\lambda \sim \delta^{\psi \nu}$. b) The correlation length $\xi$ obtained by analyzing correlation function data versus distance $\delta$ from quantum criticality. The solid line is a  fit to a power law, resulting in a quantum critical point of $\bar{\alpha}_c=-0.85(3)$ and the correlation length exponent
$\nu=2.0(2)$.}
\label{fig:fig4}
\end{figure}

An advantage of our method is that it gives direct access to the temperature dependencies
of observables. For example,  we calculate the zero-field order parameter susceptibility 
 as a function of temperature for various values of the control parameter  $\bar{\alpha}$ according to Eq.~(\ref{susc1}).  At low temperatures, the
Griffiths power law (\ref{Ordsusc0}) describes the data very well (see  Figure~\ref{fig:fig2}). The non-universal Griffiths exponent $\lambda$ can be determined from fits
in the temperature range $T=10^{-3}-1.5\times 10^{-2}$.   Figure~\ref{fig:fig4}(a) shows how $\lambda$ varies  as the distance from quantum criticality $\delta=\bar{\alpha}-\bar{\alpha}_c$ changes.  The power law $\lambda\sim \delta^{\nu\psi}$ describes the data  well  with the quantum critical point $\bar{\alpha}_c=-0.85(3)$, and exponents $\nu=2.0(2)$ 
and $\psi=0.51(2)$. Here, the number in brackets indicates the uncertainty in the last digit. These results are   consistent with the predictions of Refs.~\cite{2007_Hoyos_PRL,2009_Vojta_PRB}  and are in agreement within small errors with values found in Ref.~\cite{2008_Maestro_PRL}.

\begin{figure}[t]
\centering
\includegraphics[width=8.2cm]{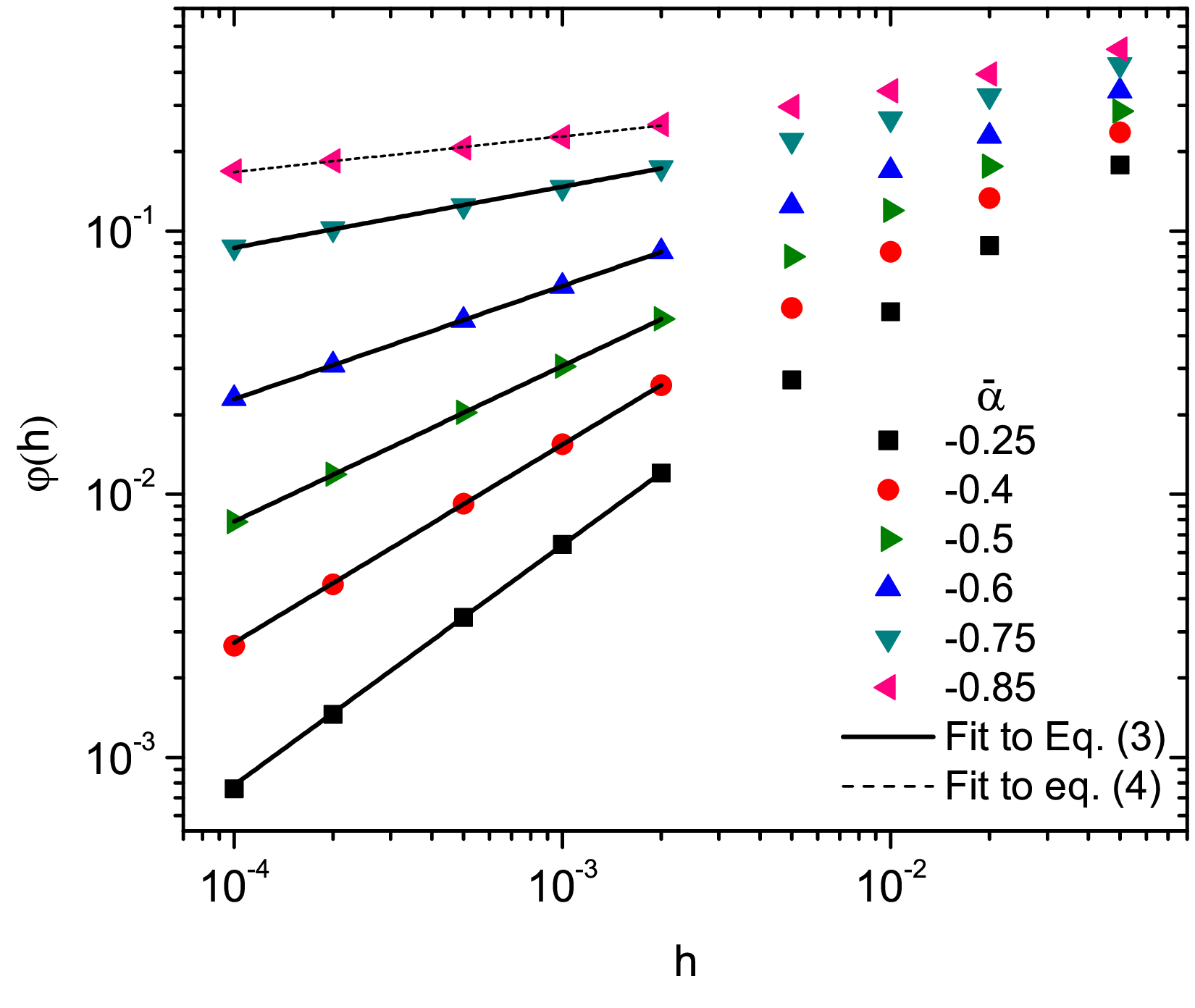}
\caption{(Color online) Order parameter $\phi$ versus  external field $h$ for various $\bar{\alpha}$. The data are averaged
over 3000 disorder configurations of system size  $L=256$.  In the field range $h=10^{-4}$  to $2\times 10^{-3}$, the   dotted and solid lines represent fits to Eq.~(\ref{Orderpcrit}) and  the Griffiths power law (\ref{Orderpgrif}), respectively.}
\label{fig:fig3}
\end{figure}

We also  compute the order parameter as a function of an external field at $T=10^{-3}$ for various  $\bar{\alpha}$ (Figure~\ref{fig:fig3}).  The off-critical data ($\delta>0$) are described by the Griffiths power law (\ref{Orderpgrif}) with an exponent $\lambda$.
At the critical point, the $\varphi(h)$ curve follows  the logarithmic dependence (\ref{Orderpcrit}) with exponents $\psi=0.51(2)$ and $\phi=1.61(2)$.
The   value for exponent $\phi$ is in agreement with the predicted one \cite{2007_Hoyos_PRL,2009_Vojta_PRB} and is consistent with the value obtained in Ref.~\cite{2008_Maestro_PRL}.
The values of the Griffiths exponent $\lambda$   match those extracted from  susceptibility data (see Figure~\ref{fig:fig4} (a)). The deviation near the critical point may be due to the fact that the correlation length becomes comparable to the 
system size and correspondingly causes  finite-size effects in the data.

\begin{figure}[t]
\centering
\includegraphics[width=9cm]{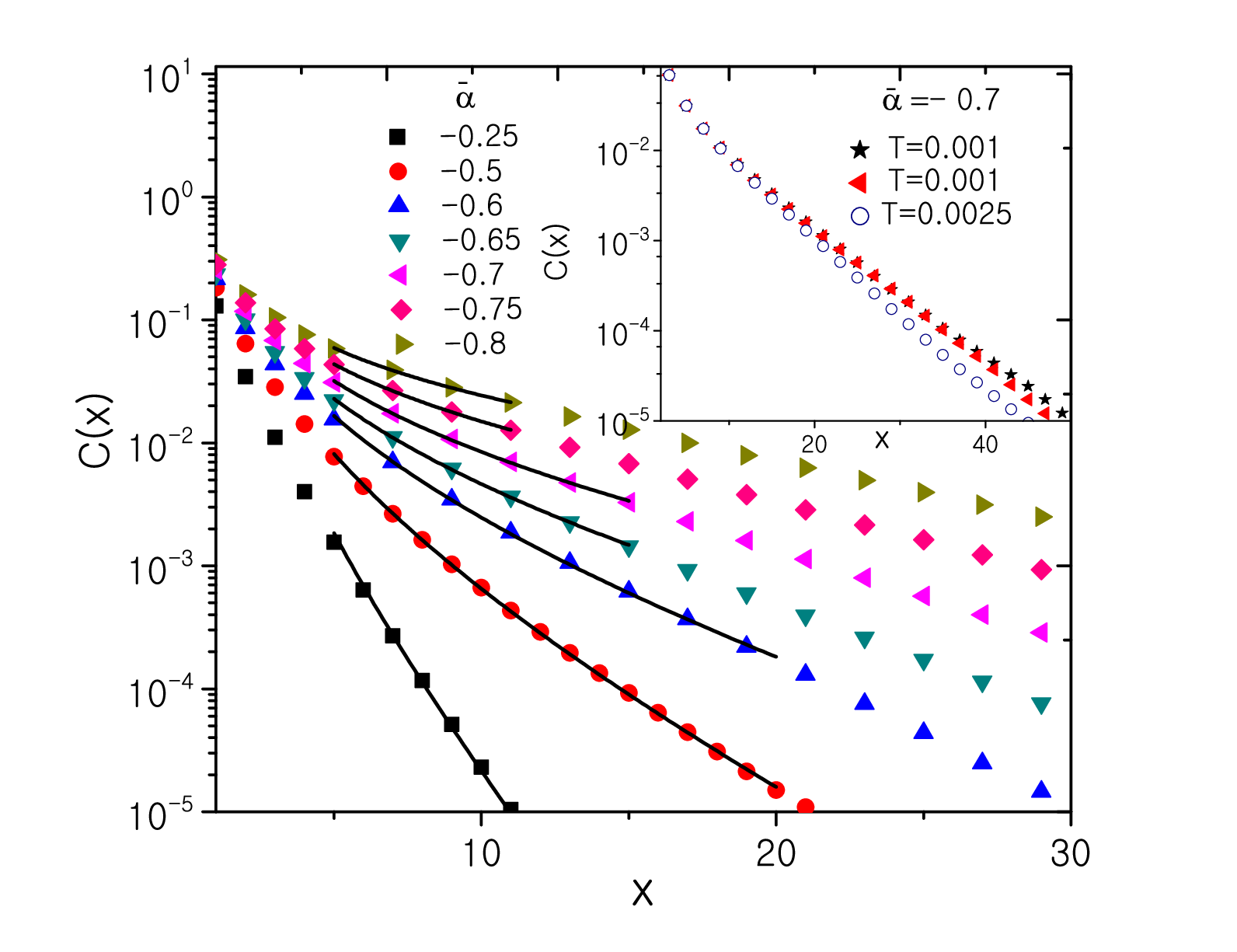}
\caption{(Color online) The equal-time correlation functions for several values of $\bar{\alpha}$. All data are
averaged over 3000 samples of size $L=1024$ at $T=10^{-3}$.  The solid lines are fits to Eq.~(\ref{corrf0}).  Inset: Deviations of correlation function
at fixed value of $\bar{\alpha}=-0.7$ due to temperature effects and statistical error of an average over disorder configurations. The data represented by circles and stars are averaged over the same 1000 disorder configurations at $T=0.0025$ and $T=10^{-3}$, respectively.  The curves represented by triangles are  averaged over   
different set of 1000 disorder configurations at $T=10^{-3}$. }\label{fig:fig1}
\end{figure}

 In addition, in the absence of an external field $h$, for system size  $L=1024,$ we compute the disorder-averaged correlation functions (\ref{corrf1})  at temperature $T=10^{-3}$
for various values of $\bar{\alpha}$ (see Figure~\ref{fig:fig1}).  The values of correlation length $\xi$ 
 can be extracted by fitting the data to Eq.~(\ref{corrf0}).
We find  good agreement of the data with Eq.~(\ref{corrf0}) for distances between $x=5$ and   some cutoff  at which the  curves start to deviate from the zero-temperature behaviors due to temperature  effects and
where curves start to become noisy because  correlations become dominated by
very rare large clusters.

Figure~\ref{fig:fig4}(b) shows how the correlation length $\xi$ changes with  distance from quantum criticality $\delta$. The data can be fitted  to the power law $\xi\sim |\delta|^{-\nu}$,
as expected \cite{1995_Fisher_PRB}. By fitting,  we  extract the critical point $\bar{\alpha}_c=-0.85(3)$ and exponent $\nu=2.0(2)$. The values of exponent $\nu$ and quantum critical point  $\bar{\alpha}_c$ are in agreement with those obtained from $\chi(T)$ and $\varphi(h)$. 

\section{Computational performance}
\label{sec:perf}
\begin{figure}[t]
\centering
\includegraphics[width=9cm]{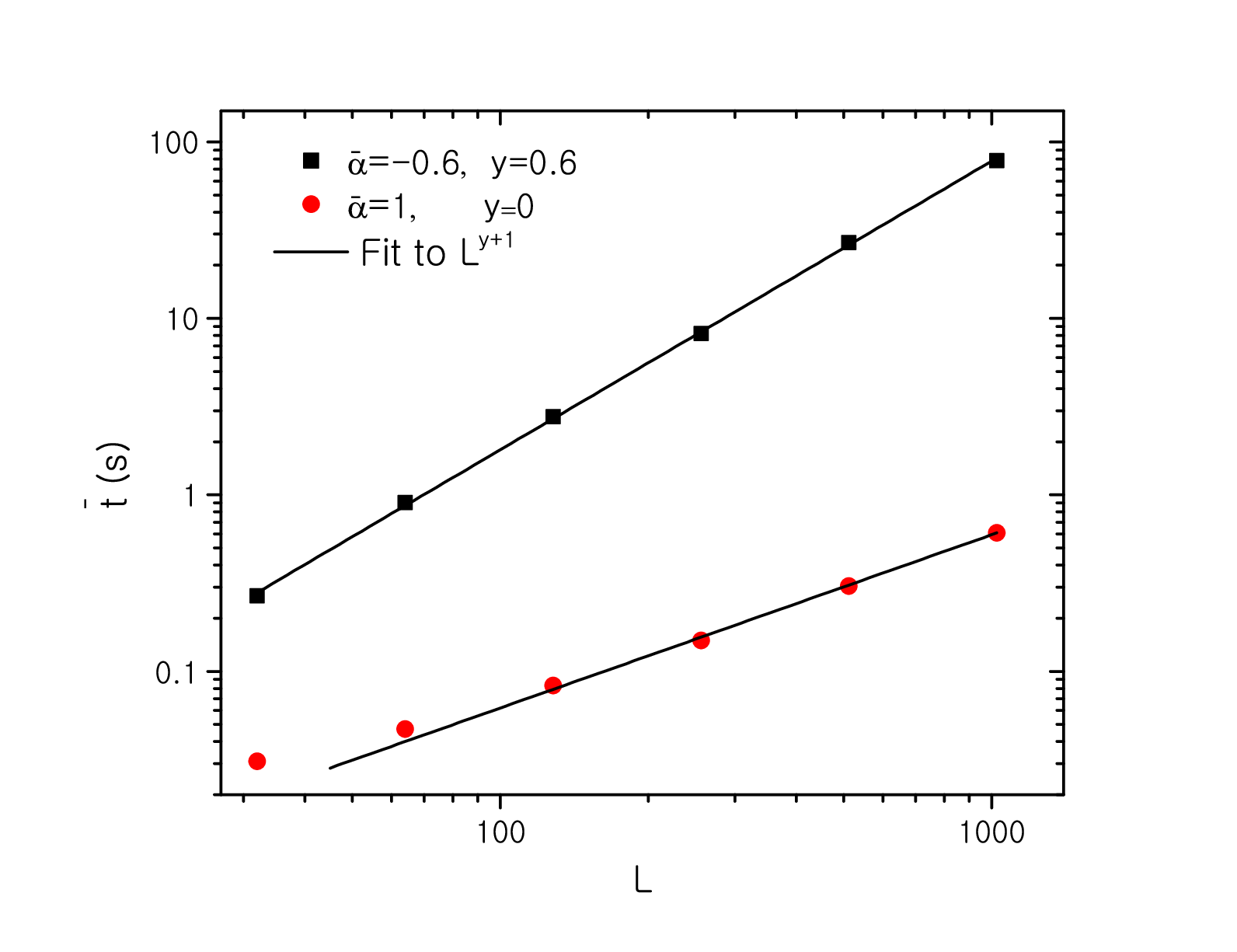}
\caption{(Color online) At the   temperature $T=10^{-3}$ and in the zero field $h=0$, execution time for a single disorder configuration $\bar{t}$  versus  system size $L$ for   $\bar{\alpha}=-0.6$ and $\bar{\alpha}=1$. All data are averaged
over 1000 disorder realizations. The solid lines represent fits to the power law. (times measured on an  Intel Core i5 CPU)}
\label{fig:fig5}
\end{figure}

In this section, we 
discuss  the execution time   of our method for solving the self-consistent Eqs.~(\ref{self12}) iteratively ({\it{i.e.}}, the time needed to get a full set of renormalized distances from criticality $r_i$). In our method, the time per iteration scales linearly with the system size $L$ in the absence of an external field because the operation count is dominated by the matrix inversion. 
Thus, the disorder-averaged execution time  $\bar{t} \sim n_{\rm{it}} L $ for a single disorder configuration, where $n_{\rm{it}}$ is the number of iterations needed
for convergence of the self-consistent Eqs.~(\ref{self12}).  The number of iterations $n_{\rm{it}}$ depends 
on the disorder configuration, it is larger for a disorder realization which has locally ordered rare regions with
smaller $\alpha$. In the  conventional  paramagnetic phase,  ${\it{i.e.}}$, for larger values of $\bar{\alpha}$  away from criticality,
locally ordered rare regions are almost absent, therefore the number of iterations $n_{\rm{it}}$ is a constant. Thus,
in the conventional  paramagnetic phase, the   execution time is expected to scale linearly with the system size, $\bar{t} \sim   L$.  Figure~\ref{fig:fig5} shows that  it indeed  scales linearly with the system size for $\bar{\alpha}=1$. In contrast, 
 in the quantum Griffiths phase, where  locally ordered rare regions are present, 
$n_{\rm{it}}$ is expected to be  large and to become larger close to criticality. If we compare two different system sizes   in the quantum Griffiths phase,  the larger system is expected to have locally ordered rare region  with higher probability.
Thus, in the quantum Griffiths phase the number of iterations $n_{\rm{it}}$ 
is expected to be a function of system size $L$, which we model as $n_{\rm{it}}\sim L^{y}$ with some non-negative exponent $y$. Therefore, in the quantum Griffiths phase the execution time does not scale linearly with the system size but it behaves as $\bar{t} \sim L^{y+1} $.
Figure~\ref{fig:fig5} shows that  for $\bar{\alpha}=-0.6$ in the quantum Griffiths phase,  the disorder averaged execution time $\bar{t}$  does not scales linearly with $L$ but behaves as   power law $\bar{t}\sim L^{y+1}$ with $y=0.6$. 
We expect the exponent $y$  to diverge as the quantum critical point is approached because the characteristic energies are exponentially small in the system size at criticality (for zero temperature and external field).  For the largest systems studied ($L=1024$), the CPU time on an intel i5 CPU
 was about 100s per disorder realizations (at $T=10^{-3}$ and $h=0$). The total numerical effort for the data presented in Sec. 5 was about 1200 CPU hours. 

Because our method performs the Matsubara sums numerically, the
effort increases with decreasing temperature  $T$.  As shown 
in Appendix \ref{app:11}, this increase is only logarithmic in $1/T$ if
we approximately combine higher Matsubara frequencies.

\section{Conclusions}
\label{sec:concl12}

In summary, we have developed an efficient numerical method for studying
quantum phase transitions in  disordered systems with $\mathcal{O}(N)$  order parameter symmetry in the large$-N$
limit. Our algorithm solves iteratively the large$-N$
self-consistent equations   for the renormalized distances
from quantum criticality
using the fast method of Ref.~\cite{1992_Meurant_SJMAA} 
for the necessary matrix inversions.  We have applied 
our method to the superconductor-metal quantum phase transition in nanowires and
 studied the critical behavior of various observables near the transition.
Our results are in agreement with strong-disorder renormalization predictions \cite{2007_Hoyos_PRL,2009_Vojta_PRB}   that the quantum phase transition is governed by 
infinite-randomness critical point accompanied by quantum Griffiths singularities.

Let us compare the performance of our method with that of the method proposed in Ref.~\cite{2008_Maestro_PRL} and outlined in
Sec.~\ref{sec:three_Adr}. The main difference is how  the sums
over the Matsubara frequencies in the self-consistent equations (\ref{rend}) are handled. The method of Ref.~\cite{2008_Maestro_PRL} 
 works at $T=0$ where the Matsubara sum becomes an integral.
 This integral is performed analytically which saves computation
 time. However, the price is a complete diagonalization  of
 the coupling matrix $M$ which is very costly ($\mathcal{O}(L^3)$
 operations per iteration). Moreover, observables at $T\neq 0$ 
 are not directly accessible. 
 
 In contrast, our method performs the Matsubara sum numerically which allows us to use the fast matrix inversion of Ref.~\cite{1992_Meurant_SJMAA}   (which needs just $\mathcal{O}(L)$
  operations per iteration) instead of a full diagonalization. Furthermore, we can calculate observables at $T\neq 0$ in contrast to  Ref.~\cite{2008_Maestro_PRL}. However,
 our effort increases with decreasing $T$.  Thus, the two methods are in some sense complementary.  The method of Ref.~\cite{2008_Maestro_PRL} is favourable for small systems when
true $T=0$ results are desired. Our method works better for larger
systems at moderately low temperatures.

We also emphasize that all our results have been obtained by converging the self-consistent equations (\ref{rend}) by means of a simple mixing scheme. Even better performance could be obtained by
combining our matrix inversion scheme with the solve-join-patch algorithm \cite{2008_Maestro_PRL} for convergence acceleration.

Our method  can be generalized to higher-dimensional problems. The self-consistent equations can be solved in the same way,   using a fast method for inverting the arising sparse matrices.  For two dimensional  systems, one could use the methods given in Refs.~\cite{2008_Li_JCP,2009_Lin_CMS} for which the cost of inversion is $\mathcal{O}(N_s^{3/2})$, where
$N_s$ is a total number of sites. We therefore expect  the cost of our method to scale  as $N^{y+3/2}_s$  or $N^{3/2}_s$ in the quantum Griffiths  and quantum paramagnetic phases, respectively. For three dimensional systems, sparse matrices can be inverted in  $\mathcal{O}(N^{2}_s)$ operations 
\cite{2009_Lin_CMS}, correspondingly the cost of our method is expected to behave as $N^{y+2}_s$  ($N_s$ is number of sites) in the quantum Griffiths phase. In the quantum paramagnetic phase it should scale as $N^{2}_s$. 

\begin{figure}[t]
\centering
\includegraphics[width=8.2cm]{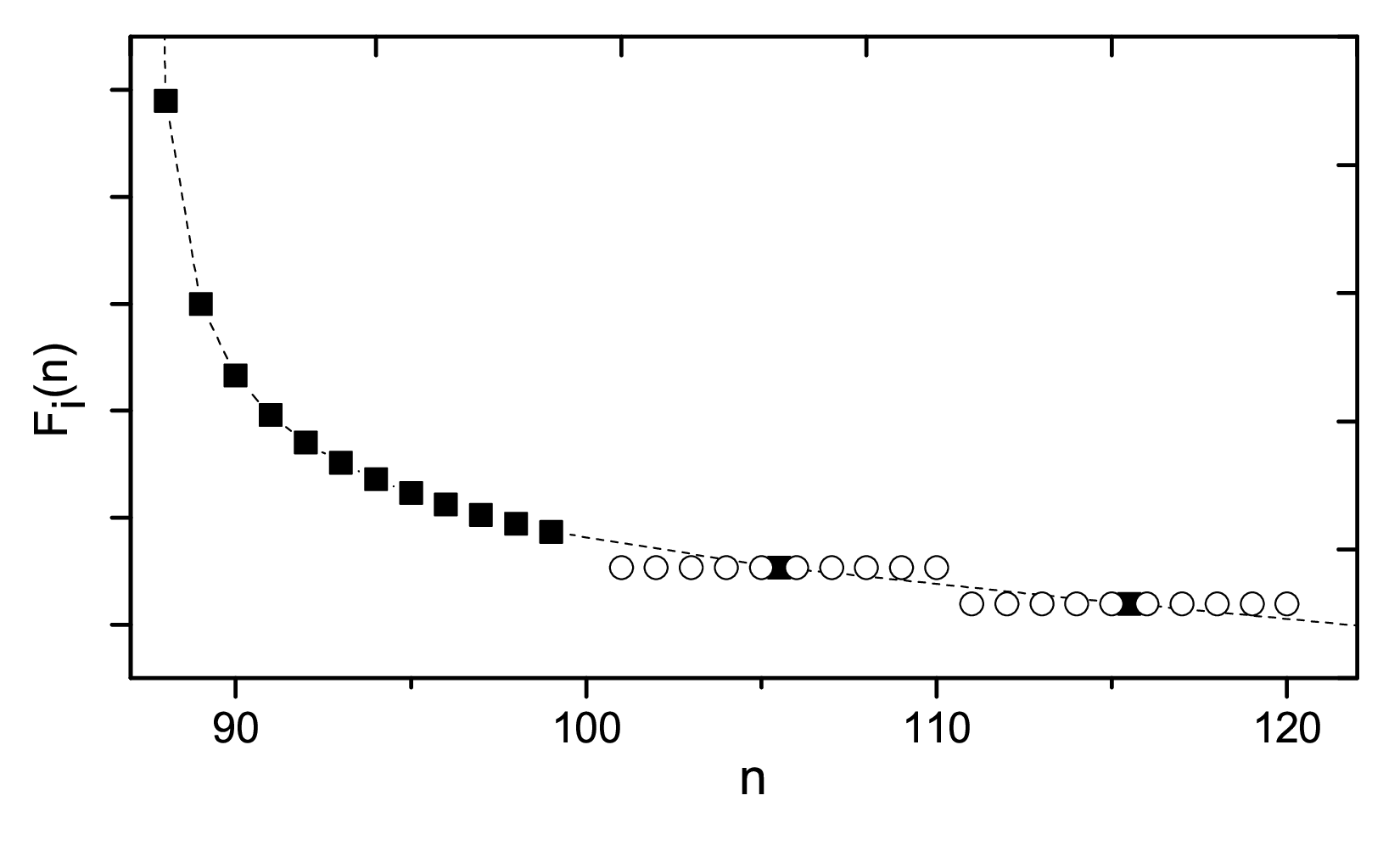}
\caption{(Color online)  Schematic of the acceleration method for the summation over the Matsubara frequencies $\omega_n$. The inverse diagonal element $F_{i}(n)$ versus $n$. The solid   squares correspond to the terms
calculated exactly by matrix inversion. The empty circles represent approximated terms. The dashed line follows  exact points.}
\label{fig:fig6}
\end{figure}

A possible application of our method in three dimensions is the disordered itinerant antiferromagnetic quantum phase transitions \cite{2007_Hoyos_PRL,2009_Vojta_PRB}. The clean transition is described 
by a Landau-Ginzburg-Wilson theory which is generalization of  the action  (\ref{action1})  to $d=3$ space dimensions   and $N=3$ order parameter components \cite{1976_Hertz_PRB,1993_Millis_PRB}. Introducing disorder leads to random mass terms as in the case of the superconductor-metal quantum phase transition in nanowires.

\section{Acknowledgements}
This work has been supported by the NSF under
Grant Nos. DMR-0906566 and DMR-1205803.
\appendix

\section{Inversion of tridiagonal matrix}
\label{app:12}
In this Appendix we sketch  the fast method for the inversion of a tridiagonal matrix  outlined in Ref.~\cite{1992_Meurant_SJMAA}. The cost of finding the diagonal elements of the
inverse matrix is  $\mathcal{O}(L)$ operations while inverting the full matrix costs $\mathcal{O}(L^2)$ operations.
The basic idea is that the inverse matrix of the tridiagonal matrix $M_{ij}$   can be represented by two sets of vectors $v_j$ and $u_j$: 
$M^{-1}_{ij}=u_iv_j$. Let  diagonal and offdiagonal elements of matrix $M_{ij}$ be $M_{ii}=a_i$  and $M_{i,i+1}=M_{i+1,i}=-b_i$, respectively. 
By combining   a UL decomposition of the linear system for $v$ and a UL decomposition of  $M_{ij}$, one can determine the set of vectors  
\begin{align} \label{inv1}
v_1=\frac{1}{d_1}\,, \hspace{0.3cm} v_i=\frac{b_2\cdots b_i}{d_1\cdots d_{i-1}d_i}\,, \hspace{0.3cm} i=2,\cdots,n\,,
\end{align}
where
\begin{align} \label{inv11}
d_n=a_n\,, \hspace{0.3cm} d_i=a_i-\frac{b^2_{i+1}}{d_{i+1}}\,, \hspace{0.3cm} i=n-1,\cdots,1\,.
\end{align}

The set of vectors $u_j$ can be found by combining  a LU decomposition of the linear system for $u$ and a  LU decomposition of  $M_{ij}$, yielding 
\begin{align} \label{inv1}
u_n=\frac{1}{\delta_n v_n}\,, \hspace{0.1cm} u_{n-i}=\frac{b_{n-i+1}\cdots b_n}{\delta_{n-i}\cdots \delta_{n}v_n}\,, \hspace{0.1cm} i=1,\cdots,n-1\,,
\end{align}
where
\begin{align} \label{inv11}
\delta_1=a_1\,, \hspace{0.3cm} \delta_i=a_i-\frac{b^2_{i}}{\delta_{i-1}}\,, \hspace{0.3cm} i=2,\cdots,n\,.
\end{align}

Finding both sets of vectors needs $\mathcal{O}(L)$ operations, consequently 
the number of operations to extract the diagonal elements $M^{-1}_{ii}=u_iv_i$ of inverse matrix scales linearly with $L$ while the cost of 
finding the full inverse matrix $M^{-1}_{ij}=u_iv_j$ is $\mathcal{O}(L^2)$.

\section{Acceleration of the frequency summation} 
\label{app:11}

 In this Appendix we propose an approach to accelerate the summation over the Matsubara frequencies in  our method.     The idea is based on the fact that the critical behaviors are dominated by  low frequencies, correspondingly    only matrices associated with low Matsubara frequencies $\omega_n$ have dominant contributions in Eq.~(\ref{self12}). 
At higher $\omega_n$, consecutive matrices change very little. Therefore, instead of finding  diagonal elements $F_i(n)$ of  $[M+2\pi Tn\mathbb{1}]^{-1}$ for each Matsubara frequencies $\omega_n$, we  invert   matrices corresponding to $n=1,...,100$ and correspondingly calculating the sum of first 100 terms  in 
 Eq.~(\ref{self12}) exactly. Then, we approximate sum of the remaining terms corresponding to $n>100$ (higher Matsubara frequencies) in the following way:  we find diagonal elements
 of $[M+2\pi Tn\mathbb{1}]^{-1}$ corresponding to the  midpoints of subintervales obtained by dividing interval $n=10^{l+1}+1, ..., 10^{l+2}$ ($l=1,...,\log_{10}(m/100)$)
into $90$ subintervales of width $10^l$.  Then, we approximate  appropriate sum in Eq.~(\ref{self12}) by summing over
 terms  calculated at midpoints multiplied  by  $10^l$. Effectively, we approximate $10^l$ terms in each subinterval
 by  values at midpoints.  Figure~\ref{fig:fig6} shows a schematic of the acceleration method. 
As a result, numerical effort scales logarithmically as $\log_{10}(1/T)$ compared with  $1/T$ scaling in the case of
 exact summation.   To check the magnitude of errors
 arising due to this approximation, we have compared observables
 calculated exactly and using  acceleration method for the system
 with size $L=256$ and  control parameter $\bar{\alpha}_c=-0.6$ at the  temperature $T=10^{-3}$. We have found that the arising relative errors are less than $0.1\%$.

%
%
\providecommand{\WileyBibTextsc}{}
\let\textsc\WileyBibTextsc
\providecommand{\othercit}{}
\providecommand{\jr}[1]{#1}
\providecommand{\etal}{~et~al.}

\end{document}